\newcount\vertou
\newcount\arxiv
\ifnum\vertou=0
\def\makeabs#1{\begin{abstract}\vspace*{4mm}#1\end{abstract}\maketitle}
\def\mypreprint#1{\preprint{#1}\vspace*{1cm}}

\def\myaddr#1{\affiliation{\vspace*{3mm}#1\vspace*{1cm}}}
\newcommand{\bim}{\begin{itemize}\vspace*{-4pt}\itemsep-2pt}
\else
\documentclass{ws-procs11x85}
\def\makeabs#1{\twocolumn[\maketitle\abstract{#1}]\baselineskip=13.07pt}
\def\mypreprint#1{}
\newcommand{\myaddr}{\address}
\newcommand{\bim}{\begin{itemize}}
\fi

\usepackage{graphicx}
\usepackage{axodraw}
\usepackage{wasysym}

\newcommand{\beq}{\begin{equation}} 
\newcommand{\eeq}{\end{equation}}

\renewcommand{\Im}{\mbox{Im}}

\def\Bbar{\,\overline{\!B}{}}  
  
\newcommand{\bit}{\begin{itemize}}
\newcommand{\ei}{\end{itemize}}
\newcommand{\bed}{\begin{description}}
\newcommand{\eed}{\end{description}}
\newcommand{\ben}{\begin{enumerate}}
\newcommand{\een}{\end{enumerate}}
\newcommand{\no}{\nonumber}
\newenvironment{Eqnarray}{\arraycolsep 0.14em\begin{eqnarray}}{\end{eqnarray}}

\renewcommand{\Im}{{\mbox{\rm Im}}}
\def\beqa{\begin{Eqnarray}}
\def\eeqa{\end{Eqnarray}}
\def\gorer{\mbox{$\Rightarrow$}}

\renewcommand{\Im}{{\cal I}m}

\def\Dbar{\overline D}

\def\Bbar{\overline B}

\def\lsim{\mathrel{\rlap{\lower4pt\hbox{\hskip1pt$\sim$}}
    \raise1pt\hbox{$<$}}} 
\def\gsim{\mathrel{\rlap{\lower4pt\hbox{\hskip1pt$\sim$}}
    \raise1pt\hbox{$>$}}}         

\begin{document}  
  
\mypreprint{\hfill\vbox{\hbox{hep-ph/0310229}\hbox{October, 2003}}}  
  

\ifnum\arxiv=1  
\title{\textnormal{\lowercase{\hfill
\vbox{\hbox{\uppercase{SLAC-PUB}-10205} 
\hbox{hep-ph/0310229}
\hbox{\uppercase{O}ctober 2003}}}}\\
\vspace*{3mm}
{\boldmath Beyond the standard model with $B$ and $K$
physics\footnote{}}}
\else
\title{\boldmath BEYOND THE STANDARD MODEL WITH $B$ AND $K$ PHYSICS} 
\fi

\ifnum\arxiv=1
\author{Yuval Grossman}
\else
\author{Y. Grossman}
\fi
\myaddr{Department of Physics,  
Technion--Israel Institute of Technology, Technion City, 32000
Haifa, Israel \footnotemark\vspace{2pt} \\ and
\vspace{2pt}\\ Stanford Linear Accelerator Center, 
  Stanford University, Stanford, CA 94309 \vspace{2pt}\\ and
\vspace{2pt}\\ 
 Santa Cruz Institute for Particle Physics, 
University of California, Santa Cruz, CA 95064
\vspace{2pt}\\ 
E-mail: yuval@slac.stanford.edu
\vspace{5pt}} 

\makeabs{In the first part of the talk the flavor physics input to 
models beyond the Standard Model is described. One specific example of
such a new physics model is given: a model with bulk fermions in one
non-factorizable extra dimension. In the second part of the talk
we discuss several observables that are sensitive to new physics. We
explain what type of new physics can produce deviations from the
Standard Model predictions in each of these observables.}
 
\ifnum\arxiv=1  
\footnotetext{$^*$Invited talk presented at the 21st International
Symposium On Lepton And Photon Interactions At High Energies (LP03)
11-16 Aug 2003, Batavia, Illinois.}
\footnotetext{$^\dagger$Permanent address.}
\else
\footnotetext{$^*$Permanent address.}
\fi 

\section{Introduction}  
 
The success of the Standard Model (SM) can be seen as a proof that it
is an effective low energy description of Nature. Yet, there are many
reasons to believe that the SM has to be extended. A partial list
includes the hierarchy problem, the strong CP problem, baryogenesis,
gauge coupling unification, the flavor puzzle, neutrino masses, and
gravity. We are therefore interested in probing the more fundamental
theory. One way to go is to search for new particles that can be
produced in yet unreached energies. Another way to look for new
physics is to search for indirect effects of heavy unknown
particles. In this talk we explain how flavor physics is used to probe
such indirect signals of physics beyond the SM.

\section{New Physics and Flavor}
In general, flavor bounds provide strong constraints on new physics
models.  This fact is called ``the new physics flavor problem''.  The
problem is actually the mismatch between the new physics scale that is
required in order to solve the hierarchy problem and the one that is
needed in order to satisfy the experimental bounds from flavor
physics.\cite{Yossi} Here we explain what is the new physics
flavor problem, discuss ways to solve it and give one
example of a model with interesting, viable, flavor structure.

\subsection{The New Physics Flavor Problem}  
In order to understand what is the new physics flavor problem let us
first recall the hierarchy problem.\cite{Martin} In order to prevent the
Higgs mass from getting a large radiative correction, new physics must
appear at a scale that is a loop factor above the weak scale
\beq \label{hpscale}
\Lambda \lsim 4 \pi m_W \sim 1 \; {\rm TeV}. 
\eeq
Here, and in what follows, $\Lambda$ represents the new physics scale.
Note that such TeV new physics can be directly probed in collider searches.

While the SM scalar sector is unnatural, its flavor sector is
impressively successful.\footnote{The flavor structure of the SM is
interesting since the quark masses and mixing angles exhibit
hierarchy.  These hierarchies are not explained within the SM, and
this fact is usually called ``the SM flavor puzzle''. This puzzle is
different from the new physics flavor problem that we are discussing
here.}
This success is linked to the fact that the SM flavor structure is
special. First, the charged current interactions are universal. (In
the mass basis, this is manifest through the unitarity of the CKM
matrix.) Second, Flavor-Changing-Neutral-Currents (FCNCs) are highly
suppressed: they are absent at the tree level and at the one loop
level they are further suppressed by the GIM mechanism.  These special
features are important in order to explain the observed pattern of
weak decays.  Thus, any extension of the SM must conserve these
successful features.

Consider a generic new physics model, that is, a model where the only
suppression of FCNCs processes is due to the large masses of the
particles that mediate them. Naturally, these masses are
of the order of the new physics scale, $\Lambda$. Flavor physics, in
particular measurements of meson mixing and CP-violation, put severe
constraints on $\Lambda$.

In order to find these bounds we take an effective field theory
approach. At the weak scale we write all the non-renormalizable
operators that are consistent with the gauge symmetry of the SM. In
particular, flavor-changing four Fermi operators of the form (the
Dirac structure is suppressed)
\beq \label{genfour}
{q_1 \bar q_2 q_3 \bar q_4 \over  \Lambda^2},
\eeq
are allowed. Here $q_i$ can be any quark flavor as long as the
electric charges of the four fields in Eq. (\ref{genfour}) sum up to
zero.\footnote{We emphasize that there is no exact symmetry that can
forbid such operators. This is in contrast to operators that violate
baryon or lepton number that can be eliminated by imposing symmetries
like $U(1)_{B-L}$ or R-parity.}
The strongest bounds are obtained from meson mixing and CP-violation
measurements:
\bit
\item 
$K$ physics: $K-\overline K$ mixing and CP-violation in $K$ decays
imply
\beq\label{fscaleK}
{s\overline d s\overline d \over \Lambda^2}
 \quad \gorer \quad \Lambda \gsim 10^4 {\mbox{~TeV}}.
\eeq
\item
$D$ physics: $D-\Dbar$ mixing implies
\beq\label{fscaleD}
{c\overline u c\overline u \over \Lambda^2}
 \quad \gorer \quad \Lambda \gsim 10^3 {\mbox{~TeV}}.
\eeq
\item
$B$ physics: $B-\Bbar$ mixing and CP-violation in $B$ decays imply
\beq\label{fscaleB}
{b\overline d b\overline d \over \Lambda^2}
 \quad \gorer \quad \Lambda \gsim 10^3 {\mbox{~TeV}}.
\eeq
\ei
Note that the bound  from kaon data is
the strongest.

There is tension between the new physics scale that is required in
order to solve the hierarchy problem, Eq. (\ref{hpscale}), and the one
that is needed in order not to contradict the flavor bounds,
Eqs. (\ref{fscaleK})--(\ref{fscaleB}).  The hierarchy problem can be
solved with new physics at a scale $\Lambda \sim 1$ TeV.  Flavor
bounds, on the other hand, require $\Lambda > 10^4$ TeV.  This tension
implies that any TeV scale new physics cannot have a generic flavor
structure. This is the new physics flavor problem.

Flavor physics has been mainly an input to model building,
not an output. The flavor predictions of any new physics model are 
not a consequence of its generic structure but rather of the special 
structure that is imposed to satisfy the severe existing flavor bounds.

\subsection{Dealing with Flavor}  

Any viable TeV new physics model has to solve the new
physics flavor problem. We now describe several ways to do so that
have been used in various models.

$(i)$ Minimal Flavor Violation (MFV) models.\cite{Buras} In such
models the new physics is flavor blind. That is, the only source of
flavor violation are the Yukawa couplings. This is not to say that
flavor violation arises only from $W$-exchange diagrams via the CKM
matrix elements. Other flavor contributions exist, but they are related
to the Yukawa interactions. Examples of such models are gauge
mediated Supersymmetry breaking models\cite{yael} and models
with universal extra dimensions.\cite{UET} In general, MFV models
predict small effects in flavor physics.

$(ii)$ Models with flavor suppression mainly in the first two
generations.  The hierarchy problem is connected mainly to the third
generation since its couplings to the Higgs field are the
largest. Flavor bounds, however, are most severe in processes that
involve only the first two generations. Therefore, one way to
ameliorate the new physics flavor problem is to keep the effective
scale of the new physics in the third generation low, while having the
effective new physics of the first two generations at a higher
scale. Examples of such models include Supersymmetric models with the
first two generations of quarks heavy\cite{heavy} and Randall-Sundrum
models with bulk quarks.\cite{huber1,huber2} In general, such models
predict large effects in the $B$ and $B_s$ systems, and smaller
effects in $K$ and $D$ mixings and decays.

$(iii)$
Flavor suppression mainly in the up sector. Since the flavor bounds
are stronger in the down sector, one way to go in order to avoid them
is to have new flavor physics mainly in the up sector.  Examples of
such models are Supersymmetric models with alignment\cite{alignment}
and models with discrete symmetries.\cite{Pakvasa} In general such
models predict large effects in charm physics and small effects in
$B$, $B_s$ and $K$ mixings and decays.

$(iv)$
Generic flavor suppression. In many models some mechanism that
suppresses flavor violation for all the quarks is implemented. Examples
of such models are Supersymmetric models with spontaneously broken
flavor symmetry\cite{LNS} and models of split fermions in flat extra
dimension.\cite{AS} In general, such models can be tested with flavor
physics.




\subsection{An Example: Bulk Quarks in the Randall-Sundrum Model}

As discussed above, there are various models that solve the new
physics flavor problem in different ways. Here we give one concrete
example: the Randall-Sundrum model with bulk
quarks\cite{huber1,huber2} which belongs to the class of models that
treat the third generation differently than the first two. Thus in
this model relatively large effects are expected in the $B$ and $B_s$
systems.

The Randall-Sundrum (RS) model solves the hierarchy problem using
extra dimensions with non-factorizable geometry.  Non-factorizable
geometry means that the four-dimensional metric depends on the 
coordinates of the extra dimensions.\cite{RS1} In the simplest 
scenario one considers a single extra 
dimension, taken to be a $S^1/Z_2$ orbifold parameterized by a 
coordinate $y=r_c\,\phi$, with $r_c$ the radius of the compact 
dimension, $-\pi\le\phi\le\pi$, and the points $(x,\phi)$ and 
$(x,-\phi)$ identified. There are two 3-branes located at 
the orbifold fixed points: a ``visible'' brane at $\phi=\pi$ 
containing the SM Higgs field, and a ``hidden'' brane at 
$\phi=0$. The solution of Einstein's equations for this geometry leads 
to the non-factorizable metric
\begin{equation}\label{metric}
   \mbox{d}s^2 = e^{-2k r_c|\phi|}\,\eta_{\mu\nu}\,
   \mbox{d}x^\mu \mbox{d}x^\nu - r_c^2\,\mbox{d}\phi^2 \,,
\end{equation}
where $x^\mu$ are the coordinates on the four-dimensional surfaces
of constant $\phi$, and the parameter $k$ is of order the fundamental
Planck scale $M$. (This solution can only be trusted if $k<M$, so the 
bulk curvature is small compared with the fundamental Planck scale.) 
The two 3-branes carry vacuum energies tuned such that 
$V_{\rm vis}=-V_{\rm hid}=-24M^3 k$, which is required to obtain a 
solution respecting four-dimensional Poincar\'e invariance. In between 
the two branes is a slice of AdS$_5$ space. 

With this setup any mass parameter $m_0$ in the fundamental theory is
promoted into an effective mass parameter which depends on the
location in the extra dimension, $m=e^{-k y}\,m_0$.  For $y=r_c \pi$
and with $k r_c\approx 12$ this mechanism produces weak scale physical
masses at the visible brane from fundamental masses and couplings of
order of the Planck scale.

\begin{figure}[t]
\centerline{
\begin{picture}(400,190)(0,-15)
\put(95,-10){\includegraphics[width=7cm]{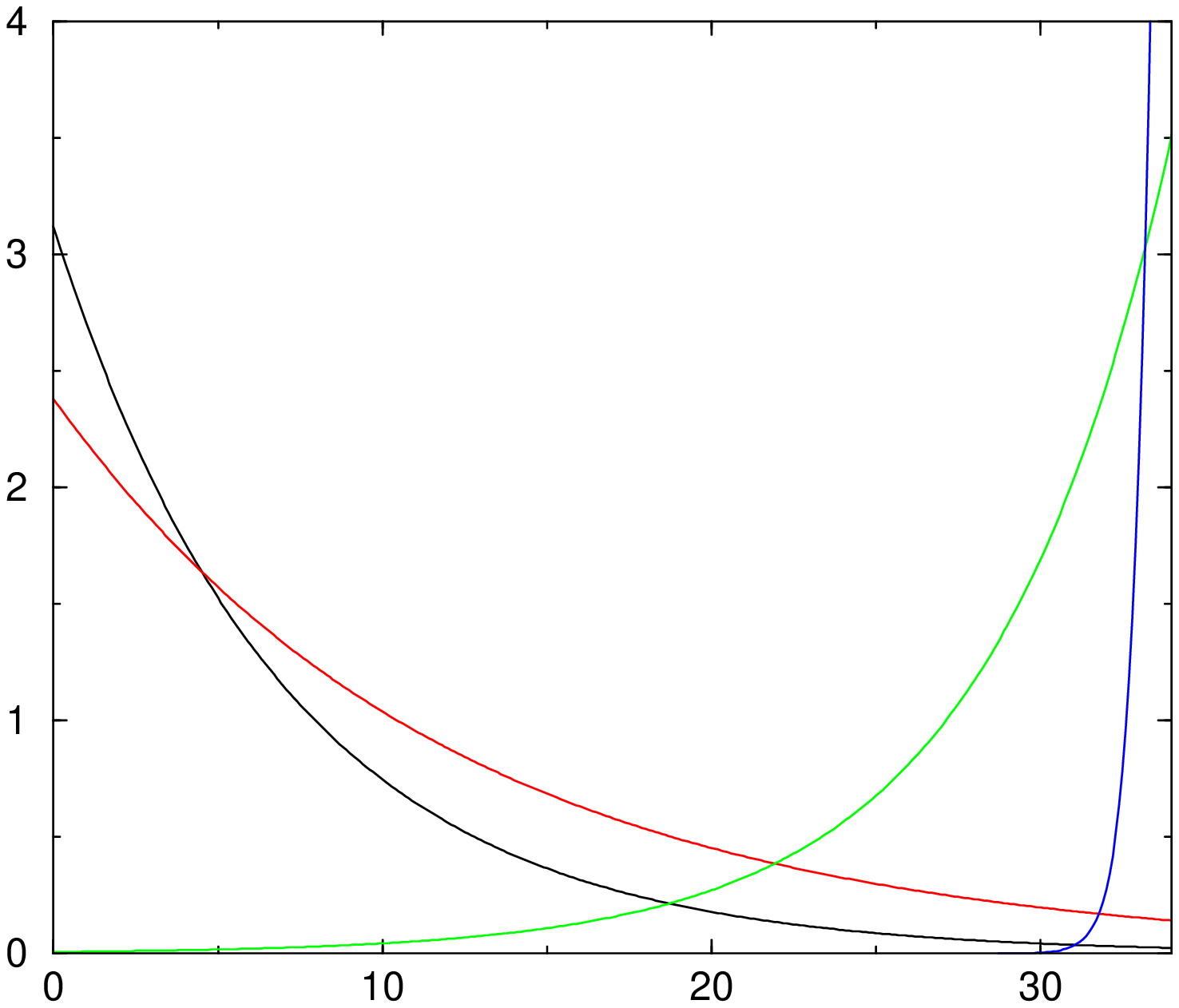}}
\put(120,95){{$q_1^{(0)}$}} 
\put(170,40){{$q_2^{(0)}$}} 
\put(250,55){{$q_3^{(0)}$}} 
\put(270,135){{$q_1^{(1)}$}} 
\put(180,-18){$ky$}
\put(246,165){{TeV-brane}} 
\put(94,165){{Planck-brane}} 
\end{picture} }
\caption[y]{An example of the shape of the fermion field wave 
functions in the RS model.\cite{huber2} $q_i^{(0)}$ are the zero mode
wave function of the $i$th generation quark doublet ($i=1,2,3$ where
$i=1$ is the lightest generation). It can be seen that the third
generation doublet is localized toward the visible brane while the
first two generation doublets are localized toward the hidden
brane. This is the reason that the effective scale of the new physics
is smaller for the third generation. 
\label{huber-fig}}
\end{figure}

The SM flavor puzzle can be solved by incorporating bulk fermions in
the RS model.\cite{Gro-Neu} Then there are several sources for new
contributions to FCNC processes. One of these new sources are
non-renormalizable operators which appear with scale of order
\beq
\Lambda \sim M \exp(-ky^f),
\eeq
where $y^f$ is the ``localization'' point of the fermion $f$. In order to
reproduce the observed quark masses and mixing
angles,\cite{huber1,huber2} heavy fermions need to have larger $y^f$,
as can be seen in Fig. \ref{huber-fig}. Thus, small effects are
expected in kaon mixing and decays and large flavor violation effects
are expected in $b$ physics.

\section{Probing New Physics with Flavor}

\begin{figure*}[t]
\begin{center}
\begin{picture}(350,180)(0,0)
\ArrowLine(20,30)(120,160)
\ArrowLine(120,160)(320,30)
\ArrowLine(320,30)(20,30)
\Text(28,100)[]{$V_{ud} V_{ub}^*$}
\Text(268,100)[]{$V_{td} V_{tb}^*$}
\Text(150,10)[]{$V_{cd} V_{cb}^*$}
\Text(121,172)[]{$(\rho,\eta)$}
\Text(122,143)[]{$\alpha$}
\Text(270,43)[]{$\beta$}
\Text(46,43)[]{$\gamma$}
\end{picture}
\end{center}
\caption[y]{The unitarity triangle. \label{UT-fig}}
\end{figure*}
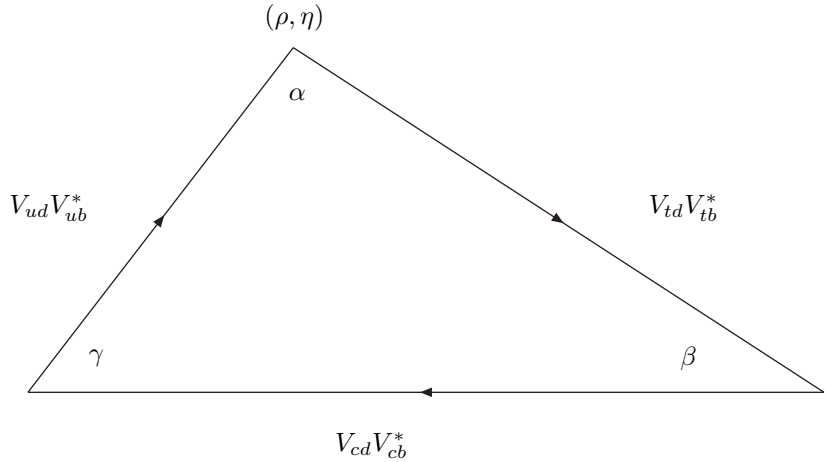

Any TeV new physics model has to deal with the flavor
bounds. Depending on the mechanism that is used to deal with flavor,
the prediction of where deviation from the SM can be expected
varies. It is important, however, that in many cases large effects are
expected.  Thus, we hope that we will be able to find such signals.

Generally, it is easier to search for new physics effects where they
are relatively large.  Namely, in processes that are suppressed in the
SM, in particular in:
\bit
\item
meson mixing,
\item
loop mediated decays, and
\item
CKM suppressed amplitudes.
\ei
It is indeed a major part in the $B$ factories' program to study such
processes. Below we give several examples for ways to search for new
physics.

Before proceeding we emphasize the following point: {\it at
present there is no significant deviation from the SM predictions in
the flavor sector}. In the following we give examples of deviations
from the SM predictions that are below the $3\sigma$ level. In
particular, we choose the following possible tests of the SM:
\bit
\item global fit,
\item $\displaystyle{a_{\rm CP}(B\to \psi K_S)}$ vs 
$\displaystyle{a_{\rm CP}(B\to \phi K_S)}$,
\item
$B\to K \pi$ decays,
\item
polarization in $B \to VV$ decays, and
\item
$K \to \pi \nu \bar \nu$ vs $B$ and $B_s$ mixing.
\ei
There are many more possible tests. Our choice of examples here is
partially biased toward cases where the present experimental ranges
deviate by more than one standard deviation from the SM
predictions. While, as emphasized above, one should not consider these
as significant indications for new physics, it should be interesting
to follow future improvements in these measurements. Furthermore, it
is an instructive exercise to think what one would learn if the
central value of these measurements turn out to be correct. As we will
see, this would not only indicate new physics, but actually probe the
nature of the new physics.

\subsection{Global Fit}

One way to test the SM is to make many measurements that determine the
sides and angles of the unitarity triangle (see Fig. \ref{UT-fig}),
namely, to over-constrain it.\cite{CKMfitter} Another way to put it is
that one tries to measure $\rho$ and $\eta$ in many possible
ways. ($\lambda$, $A$, $\rho$ and $\eta$ are the Wolfenstein
parameters.) We emphasize that this is not the only way to look for
new physics. It is just one among many possible ways to look for new
physics.

\begin{figure}[t]
\centerline{\includegraphics[width=7.8cm]{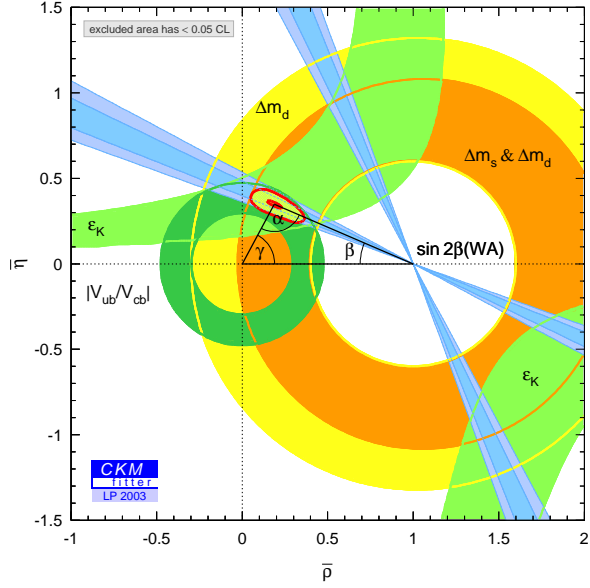}}
\caption[y]{Global fit to the unitarity triangle.\cite{CKMfitter}
The fit is based on
the measurements of $|V_{cb}|$, $|V_{ub}/V_{cb}|$, $\varepsilon_K$,
$B-\overline B$ mixing, and $a_{\rm CP}(B \to \psi K_S)$ and the bound
on $B_s$ mixing.
\label{global-fit-fig}}
\end{figure}

The global fit is done using measurements of (or bounds on)
$|V_{cb}|$, $|V_{ub}/V_{cb}|$, $\varepsilon_K$, $B-\overline B$
mixing, $B_s$ mixing, and $a_{\rm CP}(B \to \psi K_S)$. The fit is
very good, as can be seen in Fig. \ref{global-fit-fig}. Clearly, there
is no indication for new physics from the global fit. There are many
more measurements that at present have very little impact on the
fit. In the future, such measurements can be included, and then
discrepancies may show up.

\subsection{CP-Asymmetries in $b\to s \bar s s$ Modes}

The time dependent CP-asymmetry in $B$ decays into a CP eigenstate,
$f_{CP}$, is given by\cite{rev} 
\beqa\label{aCP}  
&&a_{\rm CP}(B\to f_{CP})\equiv \no\\[2pt]   
&&~~ {\Gamma(\Bbar(t) \to f_{CP}) -\Gamma( B(t) \to f_{CP}) \over  
  \Gamma(\Bbar(t) \to f_{CP}) +\Gamma( B(t) \to f_{CP})}   =
  \nonumber\\[2pt]  
&&~~ -{(1 - |\lambda|^2) \cos(\Delta m_B\, t)   
  - 2 \Im\lambda \sin(\Delta m_B\, t) \over  1 + |\lambda|^2 }  \equiv
  \nonumber\\[2pt]  
&&~~ S\, \sin(\Delta m_B\, t) - C\,  \cos(\Delta m_B\,t) .  
\eeqa  
Here $\Delta m_B \equiv m_H - m_L $ and the last line defines $S$ and
$C$. Furthermore,
\beq\label{lambdadef}  
\lambda \equiv \left({q \over p }\right)  
  \left({\bar A \over A}\right) ,  
\eeq  
where $\bar A \equiv A(\Bbar \to f_{CP})$ and
$A \equiv A(B \to f_{CP})$. 
The neutral $B$ meson mass eigenstates are defined in  
terms of flavor eigenstates as  
\beq  
|B_{L,H}\rangle = p |B \rangle \pm q |\Bbar \rangle \,.
\eeq  
In the $|\lambda| = 1$ limit, which is a very good approximation in
many cases, Eq. (\ref{aCP}) reduces to the simple form
\beq
a_{\rm CP}(B\to f_{CP}) = 
\Im\lambda\, \sin(\Delta m_B\, t)\,.
\eeq
In that case $\Im\lambda$ is just the sine of the phase between the
mixing amplitude and twice the decay amplitude.

In the SM the mixing amplitude is\footnote{Here, and in what follows,
we use the standard parameterization of the CKM matrix. The
results, of course, do not depend on the parameterization we choose.}
\beq
\arg(A_{mix})=2\beta.
\eeq
The phase of the decay amplitude depends on the decay mode.  $B \to
\psi K_S$ is mediated by the tree level quark decay $b \to c \bar c
s$ which has a real amplitude, namely,
\beq
\arg(A_{b\to c \bar cs})=0,
\eeq 
and therefore $\Im\lambda=\sin 2\beta$.  The penguin $b\to s \bar ss$
decay amplitude is also real to a good
approximation, namely,
\beq
\arg(A_{b\to s \bar ss})=0.
\eeq
We learn that also in that case $\Im\lambda=\sin 2\beta$. In
particular, the $B \to \phi K_S$, $B\to \eta' K_S$, and $B\to K^+ K^-
K_S$ are examples of decays that are dominated by the $b\to s \bar ss$
transition.  They are of particular interest since their
CP-asymmetries have been measured.  We conclude that to first
approximation the SM predicts
\beq \label{aeer}
S_{\psi K_S} = -S_{K^+ K^- K_S} =
S_{\phi K_S} = S_{\eta' K_S}.
\eeq
The theoretical uncertainties in the above predictions are
less than $O(1\%)$ for $S_{\psi K_S}$, and of $O(5\%)$ for $S_{\phi K_S}$ and
$S_{\eta' K_S}$ and $O(20\%)$ for $S_{K^+ K^-K_S}$.\cite{GLNQ}
Furthermore, for all these modes the SM predicts $|S|=\sin2\beta$. Note
that in order to violate the predictions of Eq. (\ref{aeer}), new physics
has to affect the decay amplitudes. New physics in the mixing
amplitude shifts all the modes by the same amount, leaving Eq. (\ref{aeer})
unaffected.

The data do not show a clear picture yet. Using the most recent
results,\cite{LPdata} the world averages of the asymmetries
are\footnote{We use the PDG prescription of inflating the errors when
combining measurements that are in disagreement.\cite{PDG} Simply
combining the errors there is one change in (\ref{bsssdata}), $S_{\phi
K_S}=-0.15\pm0.33$.}
\beqa \label{bsssdata}
S_{\psi K_S}&=&+0.73\pm0.05, \no\\
S_{\eta^\prime K_S}&=&+0.27\pm0.21,\no\\
S_{\phi K_S}&=&-0.15\pm0.70, \no\\
-S_{K^+K^- K_S}&=&+0.51\pm0.26^{+0.18}_{-0.00}.
\eeqa
In particular, both $S_{\phi K_S}$ and $S_{\eta^\prime K_S}$ are more
then one standard deviation away from $S_{\psi K_S}$.  (Since the
theoretical errors on $S_{K^+K^- K_S}$ are large and due to the brief
nature of this talk, we do not discuss this mode any further.)

\begin{figure*}[t]
\begin{center}
\begin{picture}(510,165)(0,-20)
\Line(0,80)(50,80)
\Line(50,80)(87,117)
\Line(85,35)(85,-15)
\Line(85,35)(134,35)
\Gluon(50,80)(85,35) 3 5
\PhotonArc(50,80)(30,45,180) 3 5
\Text(10,70)[]{$b$}
\Text(57,52)[]{$g$}
\Text(72,5)[]{$\overline q$}
\Text(85,99)[]{$s$}
\Text(120,47)[]{$q$}
\Text(40,140)[]{($P$)}
\Line(170,80)(220,80)
\Line(220,80)(257,117)
\Line(255,35)(255,-15)
\Line(255,35)(305,35)
\Photon(220,80)(254,35) 3 5
\Text(180,70)[]{$b$}
\Text(227,52)[]{$W$}
\Text(242,5)[]{$\overline u$}
\Text(255,98)[]{$u$}
\Text(290,47)[]{$s$}
\Text(210,140)[]{($T$)}
\Line(340,80)(390,80)
\Line(390,80)(427,117)
\Line(425,35)(425,-15)
\Line(425,35)(474,35)
\Photon(390,80)(425,35) 3 5
\PhotonArc(390,80)(30,45,180) 3 5
\Text(350,70)[]{$b$}
\Text(397,52)[]{$Z,\gamma$}
\Text(412,5)[]{$q$}
\Text(425,99)[]{$s$}
\Text(460,47)[]{$q$}
\Text(370,140)[]{($P_{EW}$)}
\end{picture}
\end{center}
\caption[y]{The $B \to K \pi$ amplitudes. The dominant one is
the strong penguin amplitude ($P$), and the sub-dominant ones are the
tree amplitude ($T$) and the electroweak penguin amplitude ($P_{EW}$).
\label{bkpi-fig}}
\vspace{-0.45cm}
\end{figure*}
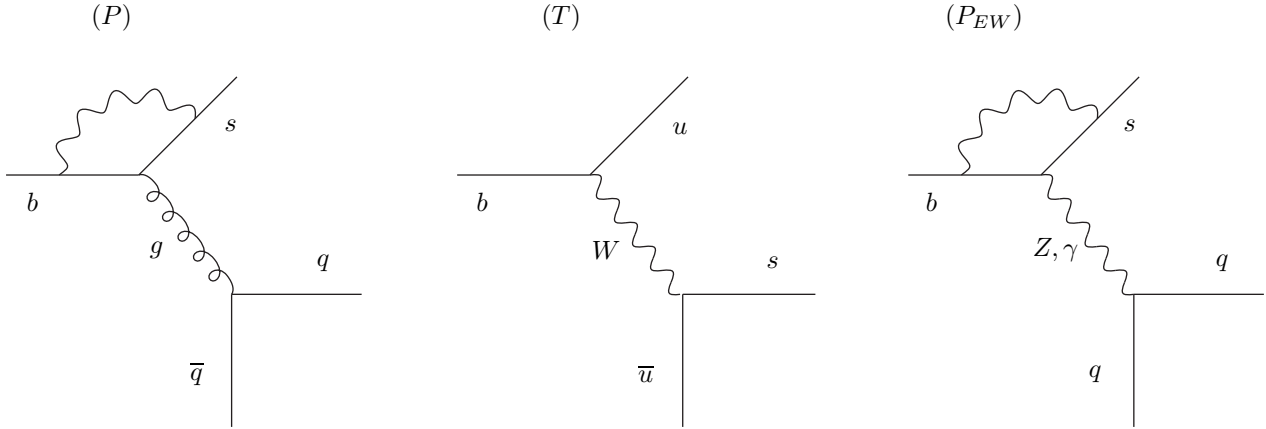

Assuming that these anomalies are confirmed in the future, we ask what
can explain them. We have to look for new physics that can generate
$S_{\psi K_S} \ne S_{\phi K_S}\ne S_{\eta^\prime K_S}$.  Since $B\to
\eta^\prime K_S$ and $B\to \phi K_S$ are one loop processes in the SM,
we expect new physics to generate large effects in the CP-asymmetries
measured in these modes. Moreover, we expect the shift from $\sin 2
\beta$ to be different in the two modes since the ratio of the SM and
new physics hadronic matrix elements is in general different.  On the
contrary, $B\to \psi K_S$ is a CKM favored tree level decay in the SM
and thus we do not expect new physics to have significant effects. We
conclude that new physics in the $b\to s\bar s s$ decay amplitude
generally gives $S_{\psi K_S} \ne S_{\phi K_S} \ne S_{\eta'
K_S}$.\cite{GrWo}

It is interesting to ask what we would learn if it turns out that
$S_{\psi K_S} \ne S_{\phi K_S}$ but $S_{\eta' K_S}$ is consistent
with $S_{\psi K_S}$. Such a situation can be the result of new parity
conserving penguin diagrams.\cite{kaganSSI,kagansuperB} To understand
this point note that $B \to \phi K_S$ is parity conserving while $B
\to \eta' K_S$ is parity violating. Thus, parity conserving new
physics in $b\to s$ penguins only affects $B \to \phi K_S$. While
generically new physics models are not parity conserving, there are
models that are approximately parity conserving. Supersymmetric $SU(2)_L
\times SU(R)\times {\rm Parity}$ models provide an example of such an
approximate parity conserving new physics
framework.\cite{kaganSSI,kagansuperB}

\subsection{$B\to K \pi$}

Consider the four $B \to K \pi$ decays and the underlying quark
transitions that mediate them:
\beqa \label{fourbkpi}
B^+ \to K^0 \pi^+ && \qquad b \to d \bar d s, \nonumber \\
B^+ \to K^+\pi^0 && \qquad b \to d \bar d s \quad {\rm or} \quad 
b \to u \bar u s,  \nonumber \\
B^0 \to K^+ \pi^- && \qquad b \to u \bar u s, \\
B^0 \to K^0\pi^0 && \qquad b \to d \bar d s \quad {\rm or} \quad 
b \to u \bar u s.  \nonumber 
\eeqa
In the SM these modes can be used to measure $\gamma$.
Moreover, there are many SM relations between these modes
that can be used to look for new physics.\cite{revbkpi}

There are three main types of diagrams that contribute to these
decays. The strong penguin diagram ($P$), the tree diagram ($T$) and
the EW penguin diagram ($P_{EW}$); see Fig. \ref{bkpi-fig}.  It is
important to understand the relative magnitudes of these
amplitudes. Due to the ratio between the strong and electroweak
coupling constants, $P\gg P_{EW}$. The relation between $P$ and $T$ is
not as simple. On the one hand, $P$ is a loop amplitude while $T$ is a
tree amplitude.  On the other hand, the CKM factors in $T$ are
$O(\lambda^2)\sim 0.05$ smaller than in $P$. Thus, it is not clear
which amplitude is dominant. Experimentally, it turns out that $P\gg
T$.  Thus, to first approximation all the four decay rates in Eq.
$(\ref{fourbkpi})$ are mediated by the strong penguin amplitude and
therefore have the same rate (up to Clebsch-Gordon coefficients).
Yet, there are corrections to this expectation due to the sub-leading
$T$ and $P_{EW}$ amplitudes.

Due to the hierarchy of amplitudes, there are many approximate
relations between the four $B\to K \pi$ decay modes. Let us consider
one particular relation, called the Lipkin sum rule.\cite{Lipkin} As
we explain below the Lipkin sum rule is interesting since the
correction to the pure $P$ limit is only second order in the small
amplitudes.

The crucial ingredient that is used in order to get useful relations is
isospin. Penguin diagrams are pure $\Delta I=0$ amplitudes, while $T$
and $P_{EW}$ have both $\Delta I=0$ and $\Delta I=1$ parts. The Lipkin
sum rule, which is based only on isospin, reads\cite{Lipkin}
\beqa
 R_L&\equiv&{2\Gamma(B^+ \to K^+ \pi^0) + 2\Gamma(B^0 \to K^0 \pi^0) 
\over 
\Gamma(B^+ \to K^0 \pi^+) + \Gamma(B^0 \to K^+ \pi^-)}  \no\\[2pt]
&=&1+ O\left(P_{EW}+T\over P\right)^2 .
\eeqa
Experimentally the ratio was found to be\cite{Fry} 
\beq
R_L=1.24 \pm 0.10.
\eeq
Using theoretical estimates\cite{Bkpith} that
\beq
{P_{EW} \over P} \sim {T \over P}\sim 0.1,
\eeq
we expect
\beq
R_L=1 + O(10^{-2}).
\eeq
We learn that the observed deviation of $R_L$ from 1 is an
$O(2\sigma)$ effect.

What can explain $R_L-1 \gg 10^{-2}$?  First, note that any new
$\Delta I=0$ amplitude cannot significantly modify the Lipkin sum rule
since it modifies only $P$. From the measurement of the four $B \to K
\pi$ decay rates we roughly know the value of $P$. This tells us that
new physics cannot modify $P$ in a significant way. What is needed in
order to explain $R_L-1 \gg 10^{-2}$ are new ``Trojan penguins'',
$P_{NP}$, which are isospin breaking ($\Delta I=1$) amplitudes. They
modify the Lipkin sum rule as follows
\beq
R_{\rm L}=1+ O\left(P_{NP}\over P\right)^2.
\eeq
In order to reproduce the observed central value
a large effect is needed, $P_{NP} \approx P/2$.\cite{Gronau-Rosner}
In many models there are strong bounds on $P_{NP}$ from $b \to s
\ell^+ \ell^-$. Leptophobic $Z'$ is an example of a viable model that
can accommodate significant Trojan penguins amplitude.\cite{GKN}

\subsection{Polarization in $B \to VV$ Decays}
Consider $B$ decays into light vectors, in particular,
\beq
B\to \rho \rho, \qquad B\to \phi K^*, \qquad B \to \rho K^*\,.
\eeq
Due to the left-handed nature of the weak interaction, in the $m_B \to
\infty$ limit we expect\cite{kagansuperB,kaganinprep}
\beq \label{SMpred}
{R_T \over R_0}= O\left({1\over m^2_B}\right),
\qquad
{R_\perp \over R_\parallel }= 1+ O \left({1\over m_B}\right)
\eeq
where $R_0$ ($R_T$, $R_\perp$, $R_\parallel$) is the longitudinal
(transverse, perpendicular, parallel) polarization fraction. Recall
that $R_T=R_\perp+R_\parallel$ and $R_0+R_T=1$.

To understand the above power counting consider for simplicity the
pure penguin $B \to \phi K^*$ decays. It is convenient to work in the
helicity  basis (${\cal A}_-$, ${\cal A}_+$ and ${\cal
A}_0$), which is related to the transversity  basis via
\beq \label{relamp}
{\cal A}_{\parallel , \perp} = {{\cal A}_+ \pm {\cal A}_-  \over \sqrt{2}},
\eeq
and the longitudinal amplitude is the same in the two bases.  We
consider the factorizable helicity amplitudes, namely, those
contributions which can be written in terms of products of decay
constants and form factors. In the SM they are proportional
to
\beqa  \label{factorizable}
&&\!\!\!\!{\cal A}_{0} \propto \frac {f_\phi m_B^3 }{m_{K^*}} \!\left[
\left(1+
\frac{m_{K^*}}{m_B} \right)\! A_1\! -\!\left(1- \frac{m_{K^*}}{m_B} \right)\! A_2
\right] \\[2pt]&&\!\!\!\! {\cal
A}_{\pm} \propto  f_\phi m_\phi m_B \left[ \left(1+
\frac{m_{K^*}}{m_B} \right)\! A_1 \!\pm \!\left(1- \frac{m_{K^*}}{m_B} \right)\! V
\right],\no
\eeqa
where terms of order $1/m_B^2$ were neglected.  The $A_{1,2}$ and $V$
are the $B\to K^*$ form factors, which are all equal in the $m_B \to
\infty$ limit\rlap{.}\,{\cite{charlesetal}} Thus, to leading-order
in $\alpha_s$\cite{BeFe}
\beq \label{FFrelations}
{A_2\over A_1} \sim {V \over A_1} = 1+{\cal
O}\left(\frac{1}{m_B}\right).
\eeq
Using Eqs. (\ref{factorizable}) and (\ref{FFrelations}) we see that
the helicity amplitudes exhibit the following
hierarchy\cite{kagansuperB,kaganinprep} 
\beq
{{\cal A}_+ \over {\cal A}_0} \sim {\cal O}\left({1\over m_B}\right),
\qquad {{\cal A}_- \over {\cal A}_0} \sim {\cal O}\left({1\over
m_B^2}\right).
\eeq
Using Eq. (\ref{relamp}) the relations in Eq. (\ref{SMpred}) immediately
follow.

An intuitive understanding of these relations can be obtained by
considering the helicities of the $q \bar q$ pair that make the vector
meson. In the valence quark approximation, when they are both
right-handed (left-handed) the vector meson has positive (negative)
helicity.  When they have opposite helicities the vector meson is
longitudinally polarized. In the $m_B \to \infty$ limit the light
quarks are ultra relativistic and their helicities are determined by
the chiralities of the weak decay operators.  Since the weak
interaction involves only left-handed $b$ decays, the three outgoing
light fermions do not have the same helicities.  For example, the
leading operator generates decays of the form
\beq
\bar b \to \bar s_R s_L \bar s_R.
\eeq
(The spectator quark does not have preferred helicity.)  Since the
$\phi$ is made from an $s$ quark and an $\bar s$ antiquark, in this
limit it has longitudinal helicity. For finite $m_B$ each helicity
flip reduces the amplitude by a factor of $1/m_B$. To get positive
helicities one spin flip, that of the $s$ quark, is required.  To get
negative helicities, spin flips of the two antiquarks are needed.

The relations in Eq. (\ref{SMpred}) receive factorizable as well as
non-factorizable corrections.  Some of these corrections have been
calculated, with the result that they do not significantly modify the
leading-order results.\cite{kaganinprep} Still, in order to get a
clearer picture, more accurate determinations of the corrections are
needed.

Observation of $R_\perp \gg R_\parallel$
would signal the presence of right-handed chirality effective
operators in $B$ decays.\cite{kaganSSI,kagansuperB} The hierarchy
between ${\cal A}_+ $ and ${\cal A}_-$ generated by the opposite
chirality operator, $\tilde{Q}_i$, (obtained from $Q_i$ via a parity
transformation) is flipped compared to the hierarchy generated by the
SM operator.  Such right-handed chirality operators lead to an
enhancement of $R_T$ and therefore can also upset the first relation
in Eq. (\ref{SMpred}).

The polarization data are as follows.\cite{Fry} The longitudinal
fraction has been measured in several modes
\beqa
{R_0(B^0\to \phi K^{*0})}&{=}&{ 0.58\pm 0.10}, \no\\ {R_0(B^+\to \phi
K^{*+})}&{=}&{ 0.46\pm 0.12}, \no\\ R_0(B^+\to \rho^0
K^{*+})&=&0.96\pm 0.16, \no\\ R_0(B^+\to \rho^+ \rho^0)&=&0.96\pm
0.07, \no\\ R_0(B^0\to \rho^+ \rho^-) &=& 0.99 \pm 0.08.
\eeqa
There is only one measurement of the perpendicular
polarization\cite{Belle-pol}
\beq
{R_\perp(B^0\to \phi K^{*0})} ={ 0.41\pm0.11}.
\eeq
Using $R_0+R_\perp+R_\parallel=1$ we extract
\beq
R_\parallel(B^0\to \phi K^{*0})=0.01 \pm 0.15.
\eeq
We see that in $B\to \rho \rho$ and $B\to K^* \rho$ the SM prediction
$R_T / R_0 \ll 1$ is confirmed, although $R_T /R_0 \gg 1/m_B^2$ remains
a possibility. Since in these modes $R_T$ is very small, the second SM
prediction, $R_\perp \approx R_\parallel$, cannot be tested yet.

The situation is different in $B\to \phi K^*$.  First, the data favor
${R_T/R_0}= O\left(1\right)$, which is not a small number. Second, one
also finds that ${R_\perp / R_\parallel} \gg1$.  Both of these results
are in disagreement with the SM predictions in Eq. (\ref{SMpred}).

It is interesting that the preliminary data indicate that the SM
predictions do not hold in $B \to \phi K^*$. This is a pure penguin $b
\to s \bar s s$ decay.  The decays where the SM predictions appear to
hold, $B\to K^* \rho$ and particularly $B\to \rho \rho$, on the other
hand, have significant tree contributions.  It is thus important to
obtain polarization measurements in other modes, especially the pure
penguin $b \to s \bar d d$ decay $B^+ \to K^{*0} \rho^+$.

With more precise polarization data it may therefore be possible to
determine whether or not there are new right-handed currents, and if
so whether or not they are only present in $b \to s \bar s s$ decays.

\subsection{$K \to \pi \nu \bar \nu$}

The $K \to \pi \nu \bar \nu$ decays are very good probes of the
unitarity triangle.\cite{Buras-Buchalla} They are dominated by the
$s\to d$ electroweak penguin amplitude with internal top quark which
is proportional to $|V_{td}|$.  Isospin and perturbative QCD can be
used to eliminate almost all the hadronic uncertainties.  One more
point that makes these modes attractive is that in many cases new
physics affects $B$ decays and $K$ decays differently.\cite{GrNi}
Then, the apparent determination of the unitarity triangle from these
different sources will be different.

Experimentally, there is only a measurement of the decay rate of the
charged mode\cite{kpnn-exp}
\beq \label{kpnn-data}
{\cal B}(K^+ \to \pi^+ \nu \bar \nu)=
(15.7^{+17.5}_{-8.2}) \times 10^{-11}.
\eeq
The SM prediction is\cite{Buras-Buchalla}
\beq \label{kpnn-SM}
{\cal B}^{\rm SM}(K^+ \to \pi^+ \nu \bar \nu)=
4.4 \times 10^{-11} \times \left[\eta^2+(1.4-\rho)^2\right].
\eeq
Using the preferred values for $\rho$ and $\eta$ (see
Fig. \ref{UT-fig}), $\rho \sim 0.15$ and $\eta \sim 0.4$, the central
value for the SM prediction is\cite{gino}
\beq \label{kpnn-SM-num}
{\cal B}^{\rm SM}(K^+ \to \pi^+ \nu \bar \nu)\approx 7.7
\times 10^{-11}.
\eeq 
We learn that the measurement [Eq. (\ref{kpnn-data})] is in agreement with
the SM prediction [Eq. (\ref{kpnn-SM-num})].

\begin{figure}[t]
\centerline{\includegraphics[width=7.8cm]{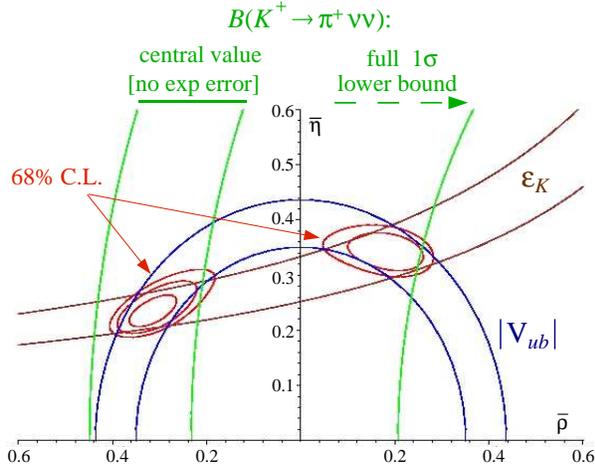}}
\caption[y]{Global fit to the unitarity triangle with the measurement
of ${\cal B}(K^+ \to \pi^+ \nu \bar \nu)$.\cite{gino} It can be seen that the
central value of the measurement is inconsistent with the unitarity
triangle extracted from the measurement of $B -\bar B$ mixing and the
bound on $B_s -\bar B_s$ mixing. 
\label{gino-fig}}
\end{figure}

It is interesting to ask what one will learn if it turns out
that the SM prediction is not confirmed by the data. Let us assume
that in the future the measurement of ${\cal B}(K^+ \to \pi^+ \nu \bar
\nu)$ will converge around its current central value.  Inspecting
Eq. (\ref{kpnn-SM}) we learn that in order to get ${\cal B}(K^+ \to
\pi^+ \nu \bar \nu)=15.7\times 10^{-11}$ we need large $\eta$
($\eta\sim 2$) or negative $\rho$. These possibilities are in conflict
with the current global fit of the unitarity triangle; see
Fig. \ref{gino-fig}. Large $\eta$ is in conflict with the measurement
of $|V_{ub}|$. Since $|V_{ub}|$ is extracted from tree level
processes, its determination is unlikely to be affected by new physics.
On the contrary, $\rho < 0$ is in conflict with the measurement of $B
-\bar B$ mixing and the bound on $B_s -\bar B_s$ mixing.  These are
loop processes, and can be modified in the presence of new physics.
We conclude that new physics in $K^+ \to \pi^+ \nu \bar \nu$ or $B-\bar
B$ mixing or $B_s - \bar B_s$ mixing can generate such a disagreement.

Higher precision in the measurement of ${\cal B}(K^+ \to \pi^+ \nu \bar
\nu)$ and a measurement of ${\cal B}(K_L \to \pi^0 \nu \bar \nu)$ are
important in order to further explore this avenue for searching for new
physics.


\section{Conclusions}
The main goal of high energy physics is to find the
theory that extends the SM into shorter distances. Flavor physics is a
very good tool for such a mission. Depending on the mechanism for
suppressing flavor-changing processes, different patterns of deviation
from the SM are expected to be found. In some cases almost no
deviations are expected, while in other we expect deviations in
specific classes of processes.  While there is no signal for
such new physics yet, there are intriguing results. More data is
needed in order to look further for fundamental physics using low
energy flavor-changing processes.

\section*{Acknowledgments}
I thank Alex Kagan, Yossi Nir, and Martin Schmaltz for helpful comments
and discussions. 
The work of YG is supported in part by a grant from the G.I.F., the
German--Israeli Foundation for Scientific Research and Development, by
the United States--Israel Binational Science Foundation through grant
No.~2000133, by the Israel Science Foundation under grant No.~237/01,
by the Department of Energy, contract DE-AC03-76SF00515 and by the
Department of Energy under grant No.~DE-FG03-92ER40689.



\ifnum\arxiv=0
\clearpage
\twocolumn[
\section*{DISCUSSION}
]
 
\begin{description}
\item[Stephen L. Olsen] (Univ. of Hawaii):
Don't long-distance effects change the SM predictions for polarization in
$B \to VV$?
\item[Yuval Grossman{\rm :}]
If $m_b$ tends to infinity, everything is short distance. The full
$1/m_b$ corrections have not been calculated yet, but they are na\"{\i}vely
expected to be of the order of $\Lambda_{QCD}/m_b$, i.e. about
10\%. The more interesting question is why the leading-order
predictions in the Standard Model hold in some $B$ decay channels, but
not in others. It is hard to believe that long-distance effect can
generate such a pattern.

\item[Rajendran Raja] (Fermilab):
If new physics is found in the $K$, $B$, and $D$ sectors, how
constraining of the new physics model will it be?
\item[Yuval Grossman{\rm :}]
If one measurement is in disagreement with the Standard Model, we
won't be able to constrain it to a specific model.  However several
indications of deviations will be model constraining, because certain
models, such as different models of Supersymmetry breaking, predict
very specific patterns of deviations from the Standard Model
predictions.
\end{description}
\fi

\end{document}